\documentclass[nohyper,12pt]{JHEP3}

\usepackage{amsmath}
\usepackage{graphicx}

\newlength{\dummysp}
\newcommand{\tr}{\mathop{{\hbox{Tr} \, }}\nolimits}

\newcommand{\stxt}[1]{\mathop{\hbox{{\scriptsize #1}}}\nolimits}

\newcommand{\half}{\frac{1}{2}}

\newcommand{\beq}{\begin{eqnarray}}
\newcommand{\eeq}{\end{eqnarray}}
\newcommand{\nnn}{ \nonumber \\ }
\newcommand{\p}{{\partial}}

\newcommand{\e}{{\epsilon}}
\newcommand{\s}{{\sigma}}
\newcommand{\vev}[1]{{\langle #1 \rangle}}

\newcommand{\ord}[1]{{{\cal O}(#1)}}
\newcommand{\gappeq}{\mathrel{\rlap {\raise.5ex\hbox{$>$}}
{\lower.5ex\hbox{$\sim$}}}}
\newcommand{\lappeq}{\mathrel{\rlap{\raise.5ex\hbox{$<$}}
{\lower.5ex\hbox{$\sim$}}}}
\newcommand{\myref}[1]{(\ref{#1})}

\newcommand{\ket}[1]{{ | #1 \rangle }}

\newcommand{\ben}{\begin{enumerate}}
\newcommand{\een}{\end{enumerate}}

\newcommand{\psib}{{\bar \psi}}

\newcommand{\bit}{\begin{itemize}}
\newcommand{\eit}{\end{itemize}}

\newcommand{\Ocal}{{\cal O}}
\newcommand{\Ncal}{{\cal N}}
\newcommand{\Ycal}{{\cal Y}}
\newcommand{\htil}{{\tilde h}}
\newcommand{\Dels}{\Delta^S}

\def\[{\left [}
\def\]{\right ]}
\def\({\left (}
\def\){\right )}

\preprint{Dec.~13, 2004 \\ hep-lat/0410041}

\title{Less naive about supersymmetric lattice quantum mechanics}

\author{Joel Giedt${}^a$, Roman Koniuk${}^b$, Erich Poppitz${}^a$,
Tzahi Yavin${}^b$ \vspace{5pt} \\
\hspace{-5pt}${}^a$Department of Physics, University of Toronto \\
60 St. George St., Toronto ON M5S 1A7 Canada \vspace{5pt} \\ 
\hspace{-5pt}${}^b$Department of Physics and Astronomy, York University \\
128 Petrie Bldg., 4700 Keele St., 
Toronto, ON M3J 1P3, Canada \vspace{5pt} \\
\hspace{-2pt}E-mail:  \email{giedt@physics.utoronto.ca}, \email{koniuk@yorku.ca},
\email{poppitz@physics.utoronto.ca}, \email{t\_yavin@yorku.ca}}

\abstract{
We explain why naive discretization results that
have appeared in [hep-lat/0006013] do not appear to yield
the desired continuum limit.  The fermion
propagator on the lattice inevitably yields a
diagram with nonvanishing UV degree $D=0$ contribution in lattice
perturbation theory, in contrast to what occurs
in the continuum.  This diagram gives a finite
correction to the boson 2-point function that
must be subtracted off in order to obtain the
perturbation series of the continuum theory,
in the limit where the lattice spacing $a$ vanishes.
Using a transfer matrix approach,
we provide a nonperturbative proof 
that this counterterm suffices to
yield the desired continuum limit.
This analysis also allows us to improve the action to $\ord{a}$.
We demonstrate by Monte
Carlo simulation that the spectrum of the continuum 
theory is well-approximated at finite but small $a$,
for both weak and strong coupling.
We contrast the above situation for the naive lattice
action to what occurs for the supersymmetric lattice
action, which preserves a discrete version of half the
supersymmetry.  There, cancellations between $D=0$
diagrams occur, obviating the need for counterterms.}


\begin{document}

\section{Introduction}

In this article we study lattice 
definitions of the simplest sort of 1d supersymmetric
quantum mechanics (SQM).\footnote{For a review, and extensive
references, see~\cite{Cooper:1994eh}.}
The continuum theory is invariant
under two independent supersymmetries ({\it susy}s) 
whose algebra is $\{ Q_1 , Q_2 \} = 2i \p_t$.
From the imaginary-time continuum action,
one can easily write down a naive lattice action.
The next task is to study the quantum continuum limit.
This exercise proves instructive as it
illustrates certain issues and subtleties in the continuum
limit of a lattice theory that contains fermions.
The lessons that are learned have relevance to
the lattice definition of supersymmetric field theories,
which are the prime motivation of this work.

The naive discretization breaks both of the
supersymmetries of the continuum theory.  
A superior lattice action has been constructed by Catterall 
and Gregory (CG) \cite{Catterall:2000rv}.  Their action
possesses one exact lattice susy that is a discrete
version of, say, $Q_1$.  However, for an interacting
theory, this lattice action cannot be made simultaneously
invariant with respect to a discrete version of $Q_2$.
The susy lattice action also follows from
a general superfield analysis \cite{Giedt:2004qs} of SQM actions
that preserve 1 nilpotent supercharge, as well
as the topological field theory approach of~\cite{Catterall:2003wd}.
CG have studied the susy lattice theory by Monte Carlo simulation.  Some
results found by CG are of interest to us in this article.
CG have computed the bosonic and fermion mass
gaps.  They find that the spectrum is degenerate,
already at finite lattice spacing $a$.
CG have compared these results to what occurs
for the case of naive discretization.  They find that
the naive discretization gives a very different answer.
Indeed, looking at their results, it would appear that
the naive discretization does not yield the
spectrum degeneracy of the continuum theory
in the $a \to 0$ limit.  In particular, the bosonic
mass gap does not approach the correct value in
this limit.

This result seems rather strange.  CG have attributed it
to large nonperturbative renormalizations.  In this
article we take a closer look at the issue.
We find that the modified fermion propagator
that occurs on the lattice gives a nonvanishing diagram
with UV degree\footnote{Our definition of UV degree
is the same as in \cite{Reisz:1987da}.  An explicit
description of UV degree will be given below.} $D=0$.  In such a diagram,
modes at the UV cutoff $a^{-1}$
do not decouple in the $a \to 0$ limit, and
their effects need to be subtracted off with a counterterm.
The susy lattice action is superior in this regard:
an additional $D=0$ diagram appears, and just cancels the
$D=0$ diagram of the naive theory.
Consequently, UV modes do decouple in the $a \to 0$ limit,
and there is no need to introduce counterterms.
These features explain the discrepant results of CG.

We now summarize the remainder of this article.
In Section \ref{prel}, we review relevant aspects
of the continuum theory.  Then we describe lattice
actions and the effect of discretization on susy.
In Section \ref{lpta}, we discuss lattice perturbation
theory and make our main points.  The modifications
to the UV behavior of perturbation theory that occur 
on the lattice have an important effect: 
one has to add a finite counterterm  
to ensure that a $D=0$ graph cancels. 
The lattice theory then has only 
graphs of negative UV degree;
Reisz's theorem \cite{Reisz:1987da} guarantees
that they approach their continuum value, 
leading to the desired supersymmetric continuum theory.
The absence of these  counterterms in \cite{Catterall:2000rv} explains why 
CG  did not obtain a susy continuum limit from the naive 
lattice action.  In Section \ref{lpta}
we also present a transfer matrix analysis, which provides
a nonperturbative proof that the counterterm suffices to
guarantee the continuum limit.  Furthermore, this
analysis allows us to derive the $\ord{a}$-improved action.

In Section \ref{swna},
we describe Monte Carlo simulations of the naive lattice
action, with and without corrections.  
We explore the effect of both (i) the necessary 
$\ord{1}$ subtraction and (ii)~the $\ord{a}$ improvement that
were derived in Section \ref{lpta}.  We find
that these corrections are sufficient to yield good
agreement with the continuum at small but nonzero $a$.
In Section \ref{concl} we draw our conclusions.

\section{Preliminaries}
\label{prel}
In this section, we review elementary aspects of the continuum
theory and set out our notations.  Certain features of the
naive lattice action and susy lattice action will be discussed.
These considerations will set the stage for our later
perturbative, transfer matrix, and Monte Carlo analyses 
of the quantum continuum limit.

\subsection{The continuum theory}
The hamiltonian of the theories that we study is of the form:
\beq
\label{HSQM}
&& H_{SQM} = \half p^2 + \half h'^2(q) - \half h''(q) [b^\dagger, b]~, \nnn
&& [q,p] = i, \qquad \{ b, b^\dagger \} = 1, \qquad
b^2 = (b^\dagger)^2 = 0~.
\label{hjre}
\eeq
Here, $q$ and $p$ are the bosonic coordinate and momentum operators,
and $b^\dagger, b$ create and destroy the fermionic state.
That is to say, a basis of states is $\ket{x,\pm}$ where 
\beq
\label{hilbertspace}
q\ket{x,\pm} = x\ket{x,\pm}, \qquad b \ket{x,-} = 0, \qquad
b^\dagger \ket{x,-} = \ket{x,+}~.
\eeq
We will refer to the function $h(q)$ that appears in
\myref{hjre} as the superpotential; there, $h'(q) = \p h(q) / \p q$, etc.  
It is well-known that the theory is invariant
under two independent susys 
whose algebra is $\{ Q_1 , Q_2 \} = 2H_{SQM} = 2i \p_t$.
These are nothing but $Q_1 = b (ip + h'(q))$ and
$Q_2 = b^\dagger (-ip + h'(q))$.

It is straightforward to write a imaginary-time
action corresponding to \myref{hjre}:
\beq
S = \int_0^\beta dt \[ \half ( {\dot x}^2 + h'^2(x)) 
+ \psib ( \p_t + h''(x) ) \psi \]~.
\label{sjre}
\eeq
Here, $\psi$ and $\psib$ are independent Grassmann fields, and $x$ is real.
All are one-component variables.  The action is invariant under
the imaginary-time continuation of the susy transformations, generated by infinitesmal
Grassmann parameters $\e_1, \e_2$:
\beq
\delta x = \e_1 \psi + \e_2 \psib~, \qquad
\delta \psib = - \e_1 (\dot x + h')~, \qquad
\delta \psi = - \e_2(\dot x - h')~.
\label{cons}
\eeq
To prove this invariance one need only make use of
periodic boundary conditions for the fields.  Equations
of motion are not involved, which is important since
they do not generally have a solution in the imaginary-time
theory.  With $\delta = \e_1 Q_1 + \e_2 Q_2$,
\beq
&& Q_1 x = \psi~, \qquad Q_1 \psib = - (\dot x + h')~, \qquad
Q_1 \psi=0~, \nnn
&& Q_2 x = \psib~, \qquad Q_2 \psib=0~, \qquad
Q_2 \psi= - (\dot x - h') ~,
\label{consy}
\eeq
defines the supercharges $Q_1, Q_2$.

In what follows it will be convenient to distinguish as $\htil$ 
the part of the superpotential that leads to interaction
terms in the action:
\beq
h = \half m x^2 + \tilde h~, \qquad
\tilde h = \sum_{n>2} \frac{g_n}{n} x^n~.
\label{seph}
\eeq
A particularly simple case that we will concentrate on
is the one studied by CG:
\beq
h = \half m x^2 + \frac{1}{4} g x^4~.
\label{siex}
\eeq

\subsection{The naive action}
Using a naive discretization of the bosons, and Wilson fermions
to avoid spectrum doublers, one obtains the following naive lattice action:
\beq
\label{naiveaction}
a^{-1} S =  \half \Delta^- x_i \Delta^- x_i 
+ \half h'_i h'_i + \psib_i (\Delta^W(r)_{ij} 
+  h''_i \delta_{ij}) \psi_j ~.
\label{naiS}
\eeq
(For convenience---here and below---we have moved the factor
of $a$ that comes from the $dt$ of the continuum to the
l.h.s.; repeated indices are summed.)  Throughout this article, we
denote the finite difference operators that we use according to:
\beq
&& \Delta^+_{ij}= \frac{1}{a} (\delta_{i+1,j} - \delta_{ij}), \qquad
\Delta^-_{ij}= \frac{1}{a} (\delta_{ij} - \delta_{i-1,j}) ,\nnn
&& \Delta^S = \half(\Delta^+ + \Delta^-), \qquad
\Delta^2 = \Delta^- \Delta^+ = \Delta^+ \Delta^- ~,
\eeq
and $\Delta^W(r)$ is the Wilson operator, defined in \myref{wop} below.
We could have chosen the forward difference operator $\Delta^+$
just as well in the bosonic part of the action; simple
identities show that for periodic 
boundary conditions
\beq
\Delta^- x_i \Delta^- x_i=\Delta^+ x_i \Delta^+ x_i~.
\label{idpm}
\eeq

As stated, we have introduced the Wilson operator
\beq
\Delta^W(r) = \Dels - \frac{ra}{2} \Delta^2
\label{wop}
\eeq
in the fermion kinetic term.
In 1d it interpolates between various choices
for the finite difference approximation to the time
derivative:
\beq
\label{wofr}
\Delta^W(r= \pm 1) = \Delta^{\mp}, \qquad \Delta_W(r=0) = \Dels ~.
\eeq
Thus, the naive choices $\Delta^+$ or $\Delta^-$ for the 
fermion kinetic term are implicitly contained in \myref{naiS}.
Taking advantage of \myref{idpm}, we obtain the naive
discretization of all time derivatives appearing in \myref{sjre}
according to $\p_t \to \Delta^-$ if $r=1$ 
and $\p_t \to \Delta^+$ if $r= -1$.
These are free of doublers, whereas $\Delta^S$ has doublers.
There are two reasons why we use the more general
operator $\Delta^W(r)$.  

First, for $d>1$, i.e.~in the field
theories that we ultimately have in mind, 
the naive choices of $\p_\mu \to \Delta_\mu^{\pm} \; (\mu=1,\ldots,d)$ always
lead to fermion doublers, as does $\Delta_\mu^S$.
By contrast, the d-dimensional Wilson operator 
$\Delta^W_{\mu}(r) = \Dels_\mu - (ra/2) \Delta^2$
does not have doublers, provided $r \not= 0$;
it does not interpolate between the naive discretizations
in $d>1$.  (Here, $\Delta^2 = \sum_\nu \Delta_\nu^+
\Delta_\nu^-$ is the d-dimensional lattice laplacian.)
Thus, we seek to understand the role of the Wilson
operator in the simpler 1d case as a foundation for
$d > 1$ lattice susy studies.

Second, the Wilson operator for $r$ such that $0 \ll |r| \ll 1$
gives a ``physical'' interpretation to UV modes,
as states of mass $\ord{ra^{-1}}$.  This is particularly
useful for our purposes in Section~\ref{lpta}.  We find a lack of decoupling
of these ``lifted doublers'' in a diagram with UV degree $D=0$ 
(cf.~Section \ref{npct}).
It is this effect that is responsible for a finite violation of susy Ward 
identities (cf.~Section \ref{fvwi})---necessitating a finite counterterm to the
naive action.  Thus, we use the Wilson operator to
illustrate the origin of the corrections coming from the
UV scale.

We choose to replace the continuum susy by a discrete approximation
$\p_t \to \Delta^+$:
\beq
&& Q_1 x_i = \psi_i, \qquad Q_1 \psib_i = - (\Delta^+ x_i + h'_i), \qquad
Q_1 \psi_i = 0 ~,\nnn
&& Q_2 x_i = \psib_i, \qquad Q_2 \psib_i =0, \qquad
Q_2 \psi_i = - (\Delta^+ x_i - h'_i) ~.
\label{nsal}
\eeq
(Other choices could be made; they do not significantly
alter our arguments.  For example, it is possible to remove
the quadratic terms in \myref{svn1} below.  However,
it can be shown that
the remaining terms will still lead to a finite violation
of the susy Ward identities.)
It is then straightforward to work out the variation of the action.
For example,
\beq
a^{-1} Q_1 S &=& -\frac{a}{2} (1+r) x_i \Delta^- \Delta^2 \psi_i 
- \frac{a}{2}(1-r) m x_i \Delta^2 \psi_i
+ \frac{ra}{2} \htil'_i \Delta^2 \psi_i \nnn
&& + \; (\Delta^S \htil'_i - \htil''_i \Delta^+ x_i) \psi_i ~.
\label{svn1}
\eeq
In each equation we have substituted
\myref{seph}. Now we note that (no sum over $i$ implied):
\beq
\Delta^S \htil'_i - \htil''_i \Delta^+ x_i =
- \frac{a}{2} \htil''_i \Delta^2 x_i + \ord{a^2}~.
\eeq
Thus we obtain that all terms on the r.h.s.~of \myref{svn1}
are suppressed by a factor of $a$.  (As is well known,
this $\ord{a}$ violation is due to the failure of
the Leibnitz rule on the lattice \cite{Dondi:1976tx}.)
A similar statement holds for $Q_2 S$.  The classical
continuum limit therefore yields
\beq
Q_A S = a {\cal Y}_A \equiv a \sum_i a Y_{A,i} 
\to \int_0^\beta dt \; a Y_A(t) \to 0~, \quad
A=1,2~.
\label{uiii}
\eeq
The susy invariance is recovered in this limit.  On the other
hand, in Section \ref{lpta} we will find that
in the quantum theory there is a finite violation
of the continuum susy Ward identities:
\beq
\vev{Q_A S} = a \vev{{\cal Y}_A}  
\sim a \cdot a^{-1} \to {\rm finite}~.
\eeq
The main point of this article is to explain why this occurs
and how to cure it.

\subsection{The susy lattice action}
Here the action is the one that appeared first
in \cite{Catterall:2000rv}.  
It preserves a discrete version of half the continuum
susy exactly.  The action was subsequently rederived
in \cite{Giedt:2004qs} using a lattice superfield
approach, and in \cite{Catterall:2003wd} using a
topological field theory approach.
It is given by (we suppress the $r$-dependence of $\Delta^W$):
\beq
a^{-1} S = \half \sum_i (\Delta^W x_i + h'_i)^2
+ \sum_{ij} \psib_i (\Delta^W_{ij} + h''_i \delta_{ij}) \psi_j~.
\label{syac}
\eeq

Comparing to the naive action~\myref{naiS}, it can
be seen that the fermionic part of the action is unchanged.  By contrast,
the bosonic part of the action differs from the naive discretization~\myref{naiS}
in a number of ways.  Due to supersymmetrization, the
Wilson operator $\Delta^W$ appears in both
the boson and fermion kinetic terms.  It follows that
at $r=0$ the bosons also have doublers; these are similarly
lifted by a Wilson mass term $m_W = m - (ra/2) \Delta^2$ for $r \not= 0$.
This degeneracy of UV modes is necessitated by the one exact lattice susy.
The lattice susy also requires derivative interaction terms that
are not present in the continuum theory.
To see this, note that
the bosonic part of the action $S_B$ may be written as:
\beq
\label{bosonic}
a^{-1} S_B &=& \half x_i [ - \Dels \Dels + (m-\frac{ra}{2}\Delta^2)^2] x_i 
+ m \htil'_i x_i + \half \htil'_i \htil'_i \nnn
&& + \htil'_i \Dels x_i - \frac{ra}{2} \htil'_i \Delta^2 x_i~.
\eeq
It is clear from (\ref{bosonic}) that 
in addition to the modified quadratic action,
there are also (in the second line) 
two sorts of interaction
vertices not present in the continuum.  The vertices
associated with $\htil'_i \Dels x_i$ come from the $h'_i \Delta^W x_i$
crossterm one gets from the square in \myref{syac}.
As pointed out in \cite{Giedt:2004qs},
these interactions break reflection positivity;
nevertheless the continuum hamiltonian and Hilbert
space are recovered after a ``conjugation'' of states
and the transfer matrix---by the operators $\exp(\pm h(q))$, 
where $q \ket{x_i} = x_i \ket{x_i}$.  It should also
be noted that whereas $\int \htil' \p x = \int \p \htil$ vanishes in the
continuum with periodic boundary conditions, on the
lattice $\htil'_i \Dels x_i$ is not purely a total derivative,
due to the failure of the Leibnitz rule.
The vertices associated with $ra \htil'_i \Delta^2 x_i$ are ``Wilson interaction
terms.''  Both of these sets of vertices have UV
degree $D=1$, so that lattice power counting
differs from that of the continuum.  We will examine this
issue in Section \ref{pcsl} below.

As mentioned above, our detailed considerations
will focus on the superpotential \myref{siex}.
In what follows, we find it convenient to set $r=1$;
this leads to various simplifications of the above formulae, which
we summarize here:
\beq
&& M_{ik} \equiv a(\Delta^-_{ik} + h''_{ik}) =
- \delta_{i-1,k} + \mu_i \delta_{ik}, \nnn
&& \mu_i = 1 + a(m + 3 g x_i^2), \qquad
\det M = - 1 + \prod_i \mu_i ~.
\label{kers}
\eeq
The matrix $M$ is the one which appears in the fermionic
part of both the naive and susy actions:  $S_F = \psib_i M_{ik} \psi$.
For $m>0$ and $g>0$, the fermion determinant
is strictly positive.  This is a happy state of affairs, given the
sign/complex-phase problems that are often experienced in
susy lattice systems; for example \cite{Giedt:2003vy}.
We note, however, that when  $h$ is a polynomial of odd degree, 
positivity will not hold.  (This happens to be
the case where it is known from the continuum that
susy is spontaneously broken.)

As first pointed out by CG, the action may be written
in the {\it Nicolai map} form:
\beq
a^{-1} S = \half \Ncal_i \Ncal_i 
+ \psib_i \frac{\p \Ncal_i}{\p x_k} \psi_k~, \qquad
\Ncal_i = \Delta^W x_i + h'_i~.
\label{nicact}
\eeq
This form makes the exact susy rather obvious:
\beq
\delta_1 x_i = \e_1 \psi_i~, \qquad \delta_1 \psi_i = 0~, \qquad
\delta_1 \psib_i = -\e_1 \Ncal_i ~.
\label{exsy}
\eeq
The index ``1'' indicates the correspondence to the
$\e_1$ part of the continuum susy \myref{consy}, the lattice 
version of $Q_1$ which has been preserved exactly on the lattice.
It is easy check that this is also a symmetry of the 
partition function measure.
This comes with the important qualification
that periodic boundary conditions are always
assumed for both bosons and fermions.  Thus our partition
function is $Z=\tr (-1)^F e^{-\beta H}$, the Witten index.
Continuum (imaginary-time) vacuum expection values are obtained
in the limit $\beta \to \infty$.

\section{Counterterm analysis}
\label{lpta}
We will show below (Section \ref{npct}) that the naive lattice action 
does not give rise to a finite lattice perturbation series,
in the technical sense that a UV degree $D=0$ proper vertex exists.
This is in contrast to the continuum ($D=0$ contributions vanish) 
and the susy lattice ($D=0$ contributions cancel---see Section \ref{pcsl}).
A local counterterm can be introduced to render the
naive lattice theory finite.\footnote{The spirit of this
analysis is quite similar to that of \cite{Golterman:1988ta}, where
a super-renomalizable supersymmetric field theory was obtained 
from a lattice theory by the treatment of counterterms.}
We determine this
counterterm and show by various means that the subtracted
theory has the correct continuum limit.  In the process,
we are able to compute the $\ord{a}$-improved action (cf.~Section \ref{trma}),
which has a faster convergence toward the continuum limit.
In Section \ref{fvwi}, we illustrate how the continuum Ward
identities are recovered in the $a \to 0$ limit, as a consequence
of the finite counterterm.

The $D=0$ diagrams that occur are just 1-loop corrections
to the boson propagator.  $D \geq 0$ primitive diagrams do not
occur at higher loops.
We will find that the $a \to 0$ results in the naive lattice case
differ from that of the continuum due to modes
at the edges of the Brillouin zone.  Thus the discrepancy can
be associated with fermion doublers that are lifted by
the Wilson mass---the contribution of the
second term in \myref{wop} at the edges of the
Brillouin zone.  This non-negligible effect
of doublers should not come as a surprise:  
the effects of lifted doublers, with $\ord{a^{-1}}$ mass,
need not decouple in a $D=0$ diagram.  Rather, we
find that they contribute a finite result.
(While a $D=0$ diagram could contribute a term 
$\sim \ln a$, we will find that this is not the case and
only an $a$-independent counterterm will be required.)

As will be shown in Section \ref{pcsl},
in the susy lattice action, an additional
$D=0$ diagram appears.
It cancels the finite correction that
appears in the naive lattice theory, alleviating the need
for a counterterm.  The additional diagram arises because
of the boson interactions associated with supersymmetrization
of the Wilson mass term.  Thus one advantage of
the susy lattice action is that it provides a lattice perturbation
theory that is finite (in the technical sense---only $D<0$ contributions
survive for all proper vertices).  By the theorem
of Reisz \cite{Reisz:1987da}, the susy lattice action therefore 
requires no counterterms to achieve the desired
continuum limit.  This explains why 
it was previously found \cite{Catterall:2000rv} that 
the susy lattice action extrapolated
well to the continuum, whereas the naive lattice action
did not.

We then show in Section \ref{trma} that the counterterm that is required for
the naive action can be seen very easily in the transfer
matrix description utilized in \cite{Giedt:2004qs}.  
Remarkably, we are able to
determine the counterterm that is required for an arbitrary
superpotential, without having to perform any loop calculations.
With a little more algebra, we are able to obtain the $\ord{a}$-improved action. 
The transfer matrix analysis has the advantage that
it is nonperturbative.  Thus we are able to conclude
that the counterterm suffices to guarantee the
continuum limit beyond perturbation theory.
This is consistent with our simulation results,
which will be discussed in Section \ref{swna}.

\subsection{Power counting for the naive lattice theory}
\label{npct}
As elsewhere, our focus here will be on the case 
of the superpotential \myref{siex}.  We will include
all quadratic terms, including the Wilson mass for the
fermion, in the free propagator.  This does not
alter the UV degree of internal fermion lines:
since $(ra/2) \Delta^2 \sim a^{-1}$ at the edges of
the Brillouin zone, the Wilson fermion propagator has the same
degree\footnote{ 
The lattice degree of divergence of a lattice Feynman 
diagram is defined as the exponent $D$ in the scaling 
$a^{-D}$ of a lattice Feynman graph at small 
lattice spacing.  In the present case of $d=1$, a factor 
$a^{-1}$ is assigned to every loop momentum integral 
(finite sum).  The small-$a$ scaling of bosonic ($a^2$) 
and fermionic ($a^1$) propagators is read off the 
expressions for lattice propagators. 
The scaling of vertices is read off from the vertex
function that appears in the Feynman rule.  For example,
none of Feynman rules for the interactions in \myref{sintn}
scale with $a$; thus they have degree $D=0$.
On the other hand, the two new interaction in \myref{slbin}
below have Feyman rules $\sim g a^{-1} \sin(2 \pi k/N) \sim a^{-1}$
and $\sim rag \cdot a^{-2} \sin^2(\pi k/N) \sim a^{-1}$
resp.  Thus they have degree $D=1$.}
$D=1$ as $\Dels \sim a^{-1}$.  The interaction
vertices can be worked out from \myref{naiS}:
\beq
a^{-1} S_{\stxt{int}}^{\stxt{naive}} = \sum_i \( mg x_i^4 + \half g^2 x_i^6
+ 3g x_i^2 \psib_i \psi_i \)~.
\label{sintn}
\eeq
We count the number of $x^4$ vertices by $V_4$,
the number of $x^6$ vertices by $V_6$, and the
number of $x^2 \psib \psi$ vertices by $\tilde V_4$.
Each of these vertices has UV degree $D=0$.
Then in the usual manner a diagram will satisfy the following
relations between the UV degree $D$, the number
of loops $L$, the number of internal boson and fermion
lines $I_B, I_F$ resp., the number of external boson and fermion
lines $E_B, E_F$ resp., and the various vertices:
\beq
D&=&L-2I_B-I_F~, \qquad
L = 1+I_B+I_F-V_4-\tilde V_4-V_6~, \nnn
E_B+2I_B&=&4V_4+6V_6+2\tilde V_4~, \qquad
E_F+2I_F = 2\tilde V_4 ~.
\label{pcnt}
\eeq

The only $D \geq 0$ solution with $L>0$ is:
\beq
D=E_F=I_B=V_4=V_6=0, \quad L=I_F=\tilde V_4=1, \quad E_B=2~, 
\label{d0vnt}
\eeq
the 1-fermion-loop correction to the boson
2-point function, diagram (a) in Figure \ref{fd1}.
In the continuum, the corresponding expression is:
\beq
\Sigma_{cont.} = 6g\int_{-\pi/a}^{\pi/a} \frac{dp}{2\pi} \frac{-i p + m}{p^2 + m^2} = 
6g\(\half + \ord{m a}\)~.
\label{cldi}
\eeq
(Note that we take the UV cutoff to be $\Lambda=\pi a^{-1}$, as
is usual when relating lattice perturbation theory to the
continuum.)  
We notice that the $D=0$ part of the diagram (i.e.,
involving the $-ip$ part of the numerator) vanishes because
it is odd w.r.t.~$p \to -p$.  For this reason, the result is
finite in the $a \to 0$ limit, rather than log-divergent
as we would naively expect based on power-counting
alone.  At the same order in perturbation
theory, a 1-scalar-loop diagram (b) of Fig.~\ref{fd1} must be added to this; it
differs only by a factor of $-2$ relative to \myref{cldi}, yielding
the net result $-3g(1 + \ord{ma})$.

\FIGURE{
\includegraphics[height=1.5in,width=5.0in,bb=100 550 400 650,clip]{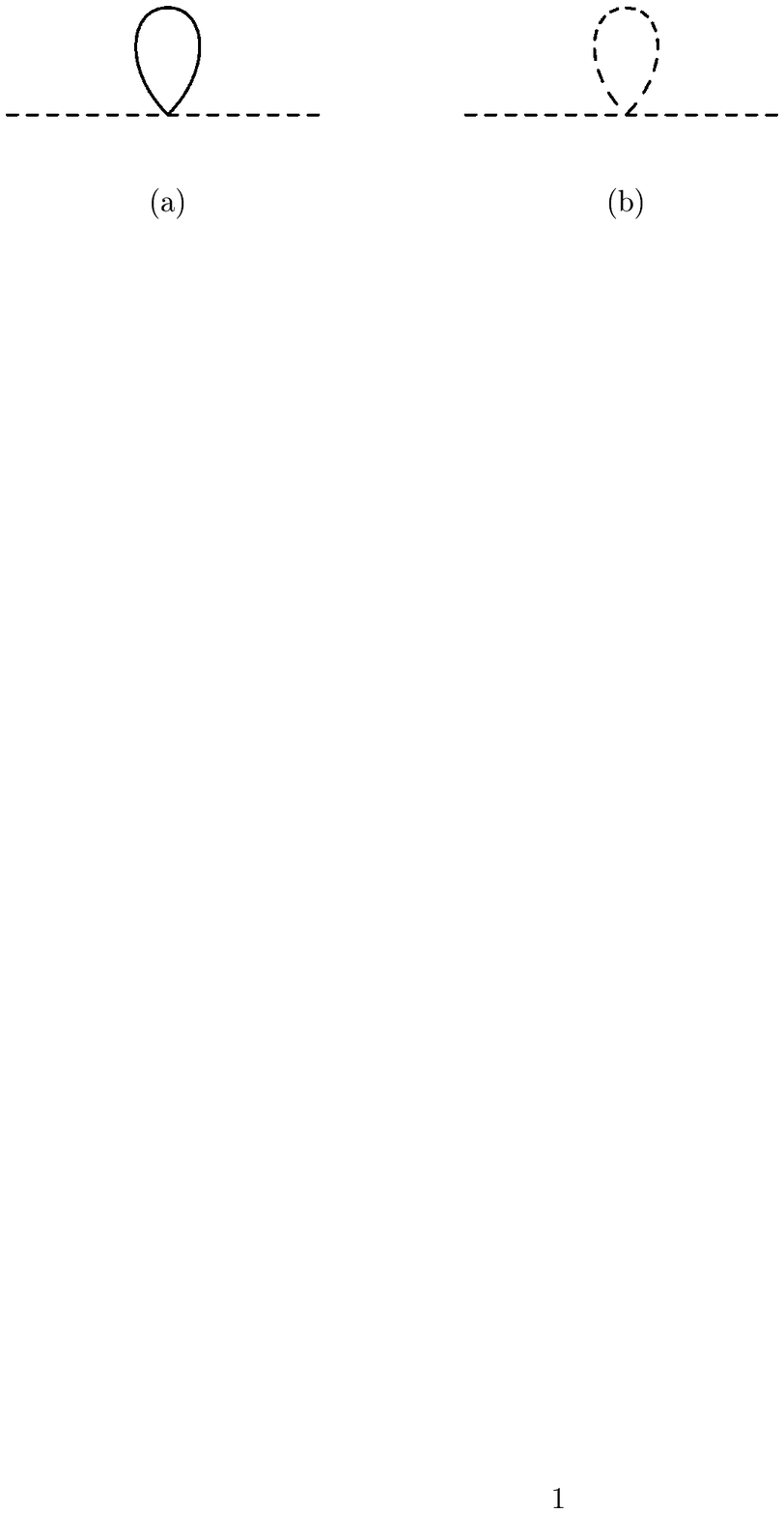}
\caption{Diagrams contributing at $\ord{g}$ to the
boson propagator.  Diagram (a) has only a $D=-1$ piece surviving
in the continuum, but on the lattice has a $D=0$ contribution coming
from fermion doublers.  In the susy lattice action, additional
interactions due to supersymmetrization of the Wilson mass term
cause diagram (b) to also acquire a $D=0$ contribution,
which just cancels that of diagram~(a). \label{fd1}}
}

On the lattice, this continuum expression
is replaced by:
\beq
\Sigma_{latt.} =
\frac{6g}{Na} \sum_{k=0}^{N-1} \frac{-i a^{-1} \sin(2 \pi k/N) 
+ m + 2r a^{-1} \sin^2(\pi k/N)}
{a^{-2} \sin^2(2 \pi k/N) + (m + 2r a^{-1} \sin^2(\pi k/N))^2} ~.
\label{latfl}
\eeq
A naive $a \to 0$ limit of this expression would
(i) set $p = 2 \pi k/(a N)$, (ii) replace $(1/Na) \sum_k
\to \int (dp/2\pi)$, and (iii) take the $a \to 0$ limit, with the 
physical momentum $p = 2 \pi k/(a N)$ fixed, on
the summand that appears in \myref{latfl}.  Then one
of course recovers the continuum fermion loop result \myref{cldi}.
However, if the $a \to 0$ limit is taken after computing
the sum, one finds that the fermion loop contribution
to the boson 2-point function is twice as big as \myref{cldi}:
\beq
\Sigma_{latt.} = 6g \( 1 + \ord{r m a} \) ~.
\label{1lia}
\eeq
This result can be understood as follows.
The $-i a^{-1} \sin(2 \pi k/N) \sim -ip$ term is odd about $k=0$
and vanishes under the sum, just as in the continuum.  However, the
Wilson mass term $2r a^{-1} \sin^2(\pi k/N)$ is even and does not.
Using $p = 2 \pi k/(a N)$, it can be written as $2r a^{-1} \sin^2(pa/2)$.
In step (iii) of the naive limit described above, we would
neglect this term.  However, in the explicit evaluation
of the sum \myref{latfl}, the Wilson mass term
becomes $\ord{a^{-1}}$ at the edges of the Brillouin
zone, where $k \approx N/2$.  In order to take
this non-negligible contribution into account,
step (iii) should be avoided,\footnote{Or, as we show below,
one can be more careful with the region where $p = \ord{a^{-1}}$.
\label{carep}} 
and the $a \to 0$ limit
must be taken after computing the sum.
Since the Wilson mass term scales like $a^{-1}$, the naive
expectations based on power counting
are correct in the lattice theory---unlike in the
continuum theory---there is a regulator dependent
contribution that must be subtracted off.  
(However, a log-divergent result is not obtained
for this particular $D=0$ diagram.)  Whereas in the continuum
we obtain \myref{cldi}, on the lattice there is an extra (finite)
contribution of $1/2$ coming from the edges of the Brillouin
zone.  The doublers do not decouple, precisely
because this is a $D=0$ diagram.

In order to match the continuum theory,
we must add a boson mass counterterm
that compensates for the effects of the
doublers in the above diagram:
\beq
\frac{6g}{Na} \sum_{k=0}^{N-1} \frac{2ra^{-1}  \sin^2(\pi k/N)}
{a^{-2} \sin^2(2 \pi k/N) + (m + 2ra^{-1}  \sin^2(\pi k/N))^2} \to 3g
\equiv \delta m^2~.
\eeq
The shifted action is then
\beq
S_c = S + \frac{a}{2} \sum_i \[ 3gx_i^2 + {\rm const.} \]
= S + \frac{a}{2} \sum_i h''_i ~,
\label{jiil}
\eeq
where we have added an overall constant ($m/2$) to obtain
the appealing form of the far r.h.s.  It will turn
out that this constant appears automatically in our
transfer matrix analysis of Section~\ref{trma} below.
We have only shown that a mass counterterm is required
for the special case of $h$ studied here, eq.~\myref{siex}.  The expression
\myref{jiil} would, for a more general $h$, imply interaction
counterterms as well.  In Section \ref{trma}
we will show that the counterterm $h''/2$ is the correct choice 
in the general case.

We now give an intuitive understanding of why a finite
counterterm is needed.  Our discussion will also give the
more careful treatment of the region where
$p = \ord{a^{-1}}$, mentioned in Footnote~\ref{carep},
and will exploit the flexibility allowed by
our use of the Wilson operator \myref{wop}.  Consider the case 
of $r$ and $a$ such that $r \ll 1$
but $m \ll r a^{-1}$.  Then the sum \myref{latfl} is well
approximated as follows.  The mass of the (lifted) doubler mode
is just $M=2ra^{-1} + m$, as can be seen by a change of
variables to $p'=p-\pi a^{-1}$, and looking near small $p'$.
Thus for $M \ll a^{-1}$, the doubler contributes like
a particle of mass $M \gg m$.  It follows that in this
limit a good approximation to the integral that
appears in \myref{1lia} is just
\beq
\int_{-\pi/a}^{\pi/a} \frac{dp}{2\pi} \frac{m}{p^2 + m^2}
+ \int_{-\pi/a}^{\pi/a} \frac{dp'}{2\pi} \frac{M}{p'^2 + M^2}
= \frac{1}{\pi} \[ \tan^{-1} \frac{\pi}{2am} + \tan^{-1} \frac{\pi}{2aM} \] ~.
\label{skker}
\eeq
To $\ord{r}$, each term on the r.h.s.~of \myref{skker} yields $1/2$.
Thus as the UV cutoff $a^{-1}$ is sent
to infinity, the contribution of the doublers does not
decouple.  For $D \geq 0$ diagrams this lack of decoupling
is usually associated with a divergence.  Here, the lack of
decoupling in a $D=0$ diagram is instead only associated with
a finite, $a$-independent contribution coming from the regulator.

Finally, we note that our result is consistent
with the power-counting theorem of Reisz~\cite{Reisz:1987da}.  In essence,
his theorem states that the naive continuum limit
can be taken on the integrand of diagrams---step (iii)
of the naive limit mentioned above---provided the UV
degree of the diagram satisfies $D<0$.  That is, for $D<0$
diagrams, the $a \to 0$ limit and the loop momenta integrals commute.
However, this theorem does not apply to a $D=0$ diagram,
regardless of whether or not it is finite (in the sense
of being $a$-independent versus $\ln a$ dependent).

\subsection{Transfer matrix analysis of the naive and improved lattice actions}
\label{trma}
The  partition function $Z_{naive}$ of the naive lattice action (\ref{naiveaction}), 
with $r=1$, can be represented as the trace 
of the $N$-th power of the transfer matrix operator, denoted here by $T[0,0]$,
over the Hilbert space (\ref{hilbertspace}), weighted
by $(-1)^F$ to account for periodic boundary conditions
of fermions.\footnote{The case of $r \not= 1$ can also be treated, but it is more complicated
because the transfer matrix must have row and column indices labeled by pairs
of sites.}  Thus, as we show in some detail in the appendix:
\beq
\label{Tmatrix0}
Z_{naive} = {\rm Tr} (-1)^F\; T[0,0]^N~,
\eeq
where $T[0,0] = T[k = 0, \ell = 0]$  can be read off eqn.~(\ref{Tmatrix}) below.
In anticipation of counterterms,
we generalize the naive action by functions $k(x)$ and $\ell(x)$:
\beq
\label{cterms1}
h'^2(x) \to h'^2(x) + k(x)~, \qquad
h''(x) \psib \psi \to (h''(x) + \ell(x)) \psib \psi ~,
\label{nacts}
\eeq
and denote the corresponding ``deformed'' transfer matrix as $T[k,\ell]$, given by the following operator:
\beq
\label{Tmatrix}
T[k, \ell] ={\cal{N}}(a)   \int\limits_{-\infty}^{\infty} dz 
\exp \[ -\frac{z^2}{2a} - \frac{a}{2} h'^2(q)
- \frac{a}{2} k(q) \] e^{izp} \[1 + a(h''(q) + \ell(q)) b^\dagger b \] ~,
\eeq
where ${\cal{N}}(a)$ is an inessential $a$-dependent normalization constant. (In the appendix we show that the naive lattice partition function supplemented by the counterterms (\ref{cterms1}) can be written as ${\rm Tr} (-1)^F\; T[k,\ell]^N$.)

For small lattice spacing, one can  
use a saddle point approximation to evaluate the $z$-integral in (\ref{Tmatrix}). Then, 
it follows that the operator $T[k,\ell]$   of (\ref{Tmatrix})  can, for small $a$,   be written as:
\beq
\label{Tmatrixsmalla}
T[k,\ell] = \; e^{a f(q)}  \; e^{- a H[k,\ell]} \; e^{-a f(q)}~, 
\eeq
where  $f$ is given by:
\beq
\label{ef}
f(q) &=& -\frac{1}{4}(h'^2(q) + k(q)) 
- \frac{1}{2} (h''(q)+\ell(q)) b^\dagger b~,
\eeq
and the hamiltonian $H[k,\ell]$ is:
\beq
\label{Thamiltonian}
H[k,\ell] &=& \half p^2 + \half h(q)'^2 
- ( h''(q) + \ell(q))\(1 - \half a (h''(q) + \ell(q))\)
\half [ b^\dagger , b] 
\nnn
&& + \half k(q)
- \half( h''(q) + \ell(q))\(1 - \half a (h''(q) + \ell(q))\)  + \ord{a^2}~.
\eeq 
Two important observations immediately follow from (\ref{Tmatrixsmalla}-\ref{Thamiltonian}):
\begin{enumerate}
\item 
The $T[k,\ell]$ operator at small lattice spacing is equivalent, 
upon conjugation by $e^{a f}$, to $\exp(- a H[k,\ell])$.
When inserted in the trace, the factors $e^{\pm a f}$ in (\ref{Tmatrixsmalla}) do not 
affect the partition function.  (Actually, this
conjugation can be avoided if one writes $T[k,\ell]$ in the
standard form $e^{-\frac{a}{2}V(q,b^\dagger b)} e^{-\frac{a}{2}p^2} 
e^{-\frac{a}{2}V(q,b^\dagger b)}$,
with an appropriate choice of $V$.) 
Thus, the general partition function 
$Z[k,\ell] \equiv {\rm Tr} (-1)^F T[k,\ell]^N$ approaches 
${\rm Tr} (-)^F e^{-\beta H[k,\ell]}$ in the small-$a$, $\beta$-fixed limit,
where $\beta \equiv Na$.
\item 
The hamiltonian $H[0,0]$---corresponding to the naive 
lattice action   (\ref{naiveaction})---does not, 
even as $a \rightarrow 0$, approach the hamiltonian $H_{SQM}$ (\ref{HSQM}) 
of SQM.  Rather, as eq.~(\ref{Thamiltonian}) 
shows, $H[0,0] = H_{SQM} - \half h''(q) + \ord{a}$.
\end{enumerate}
The second observation above shows that in order for the continuum limit of the 
naive lattice partition function to correspond  to the path integral of 
SQM, the naive lattice action has to be 
supplemented by a finite counterterm: one must take
$k(q) = h''(q)$ and $l(q) = 0$ in the transfer matrix
(up to $\ord{a}$ terms). 
This counterterm, with $h$ of eq.~(\ref{siex}), is precisely equal 
to the counterterm calculated in 
perturbation theory in the previous section, eq.~(\ref{jiil}).

Moreover, eq.~(\ref{Thamiltonian}) allows us to  go  
further and demand that the correspondence to $H_{SQM}$ is 
not corrected also at $\ord{a}$.  One chooses $k$ and $\ell$  
such that $H[k,\ell] = H_{SQM} + \ord{a^2}$, which implies
\beq
\label{improvedconditions}
k-( h'' + \ell)\(1 - \half a (h'' + \ell)\) &=& \ord{a^2}~, \nnn
h''-( h'' + \ell)\(1 - \half a (h'' + \ell)\) &=& \ord{a^2}~.
\eeq
Thus, the $\ord{a}$-improved naive lattice action should have a 
transfer matrix $T[k,\ell]$ with  
\beq
k(q) = h''(q) + \ord{a^2}~, \qquad \ell(q) = \half a h''^2(q) + \ord{a^2}~.
\eeq
The $\ord{a}$-improved action is therefore (recall we set $r=1$):
\beq
\label{improvedaction}
a^{-1} S_{ca} = \half \sum_i \( \Delta^- x_i \Delta^- x_i 
+ h'_i h'_i + h''_i \) 
+ \sum_{ij} \psib_i (\Delta^-_{ij} 
+  h''_i \delta_{ij} + \half a h''^2_i \delta_{ij}) \psi_j ~.
\label{impS} 
\eeq

The above analysis provides a complete, nonperturbative
proof that the $\ord{1}$ correction \myref{jiil} to the bosonic part of the
action suffices to guarantee the desired continuum limit.
The $\ord{a}$ improvement, which only corrects the
fermionic part of the action, will be shown in our
simulation results of Section \ref{swna} to yield very good
results in the $a \to 0$ limit.

\subsection{Susy Ward identities in the naive lattice theory}
\label{fvwi}
It is easy to show that the following
Ward identities hold on the lattice:
\beq
\vev{Q_A \Ocal} = \vev{(Q_A S) \Ocal}
= a \vev{\Ycal_A \Ocal}~, \quad A=1,2 ~.
\label{jues}
\eeq
Here $\Ycal_A$ are the lattice operators defined by \myref{uiii},
whose explicit form can be obtained from \myref{svn1}
and a similar expression for $Q_2 S$.
A violation of the continuum Ward identity $\vev{Q_1 \Ocal}=0$
can only arise if the difference operators in $\Ycal_1$ generate
a UV-divergent result, $\vev{\Ycal_A \Ocal} = \ord{a^{-n}}, \; n>0$.
We now show that this indeed occurs in the uncorrected theory.
This is in spite of the fact that the lattice perturbation
series has only a $D=0$ diagram, as shown in Section \ref{npct}.
The point is that, as can be seen from \myref{svn1}, 
$\Ycal_1$ has vertices that have positive
UV degree, whereas those that occur in the 
action \myref{sintn} have vanishing UV degree.
We will also illustrate how this difficulty is cured by the counterterm
\myref{jiil}.  

We note from \myref{svn1} that each term in
$\Ycal_1$ contains finite difference
operators.  When these operators act on would-be fermion doubler modes
(i.e., doublers when $r=0$), they give positive UV degrees.
We will study the anomalous Ward identity associated with the operator
$x_i \psib_j$, since it is directly related to
the lack of spectrum degeneracy in the unsubtracted
naive case.  
Actually, we must be careful to restrict to the physical
modes that are relevant to the ``long-distance'' physics.  Operators
at fixed time could easily be obtained by blocking the lattice fields,
but it is simpler to just analyze the Ward identies in momentum space;
i.e., using the Fourier transformed variables $\tilde x_k, \tilde \psib_k$.
Thus we will consider the Ward identities associated with the
operator $\Ocal_k = \tilde x_k \tilde \psib_k$ subject to $k \ll N/2 \mod N$.
We will find that whereas $\vev{Q_1 \Ocal_k}$ does not vanish
in the continuum limit for the naive action, once the counterterm
in introduced, the finite violation is subtracted off and the
Ward identity is recovered in the $a \to 0$ limit.

Rather than study this cancellation directly, we will
exploit the r.h.s.~of the lattice Ward identity \myref{jues}.  We break
$Q_1 S$ into operators that generate 2-point and 4-point vertices:
\beq
Q_1 S = (Q_1 S)_{(2)} + (Q_1 S)_{(4)} 
\equiv a \Ycal_{1,(2)} + a \Ycal_{1,(4)}~.
\label{ipor}
\eeq
The operator $\Ycal_{1,(2)}$ gives a $D=3$ vertex
and $\Ycal_{1,(4)}$ gives a $D=2$ vertex.
The leading order diagram associated with the latter vertex
is shown in Fig.~\ref{fd2}(a).  The leading $D=2$
piece vanishes.  The surviving $D=1$ part
yields an $\ord{a^{-1}}$ result that just cancels the factor of
$a$ in \myref{ipor}, leading to a finite result of 
$3g(1 + \ord{ma})$.  The 2-point vertex also contributes
at $\ord{g}$, through the diagram of Fig.~\ref{fd2}(c).
Here, the $3g x^2 \psib \psi$ vertex of the action \myref{sintn}
is involved.  The diagram has UV degree $D=1$,
again canceling the $a$ in \myref{ipor}, yielding
a finite result of $-6g(1 + \ord{ma})$.  
The net result is finite and nonzero, which is consistent
with the lack of spectrum degeneracy
observed in \cite{Catterall:2000rv}.  The continuum Ward identity
$\vev{Q_1 \Ocal_k}=0$ is not recovered in the $a \to 0$ limit,
for the uncorrected naive action~\myref{naiS}.

However, once the counterterm $(3/2)gx^2$ is added to the
action,
\beq
Q_1 S \to Q_1 S + a \sum_i 3g x_i \psi_i ~.
\eeq
This contributes a 2-point $D=0$ vertex that is not suppressed by $a$,
and is indicated by the tree-level diagram in Fig.~\ref{fd2}(b).
This of course yields $3g$, canceling the finite violation
that arises from the other two diagrams.
Thus in the $a \to 0$ limit, we obtain to $\ord{g}$ the continuum
result $\vev{Q_1 \Ocal_k}=0$.  We have verified
that similar cancellations occur at $\ord{g^2}$.  

Of course, it is guaranteed that the cancellations will occur to all orders
in perturbation theory, since the counterterm has ensured that
we will recover the continuum perturbation series, which satisfies
the susy Ward identities to all orders---for 
any operator $\Ocal$, not just the one that we have considered here.
Furthermore, our transfer
matrix analysis provides a nonperturbative proof that the continuum
hamiltonian is obtained in the quantum continuum limit.  
We necessarily recover the continuum result $\vev{Q_A \Ocal}=0$
in the corrected lattice theory when $a \to 0$.

\FIGURE{
\includegraphics[height=1.4in,width=5.5in,bb=50 550 450 650,clip]{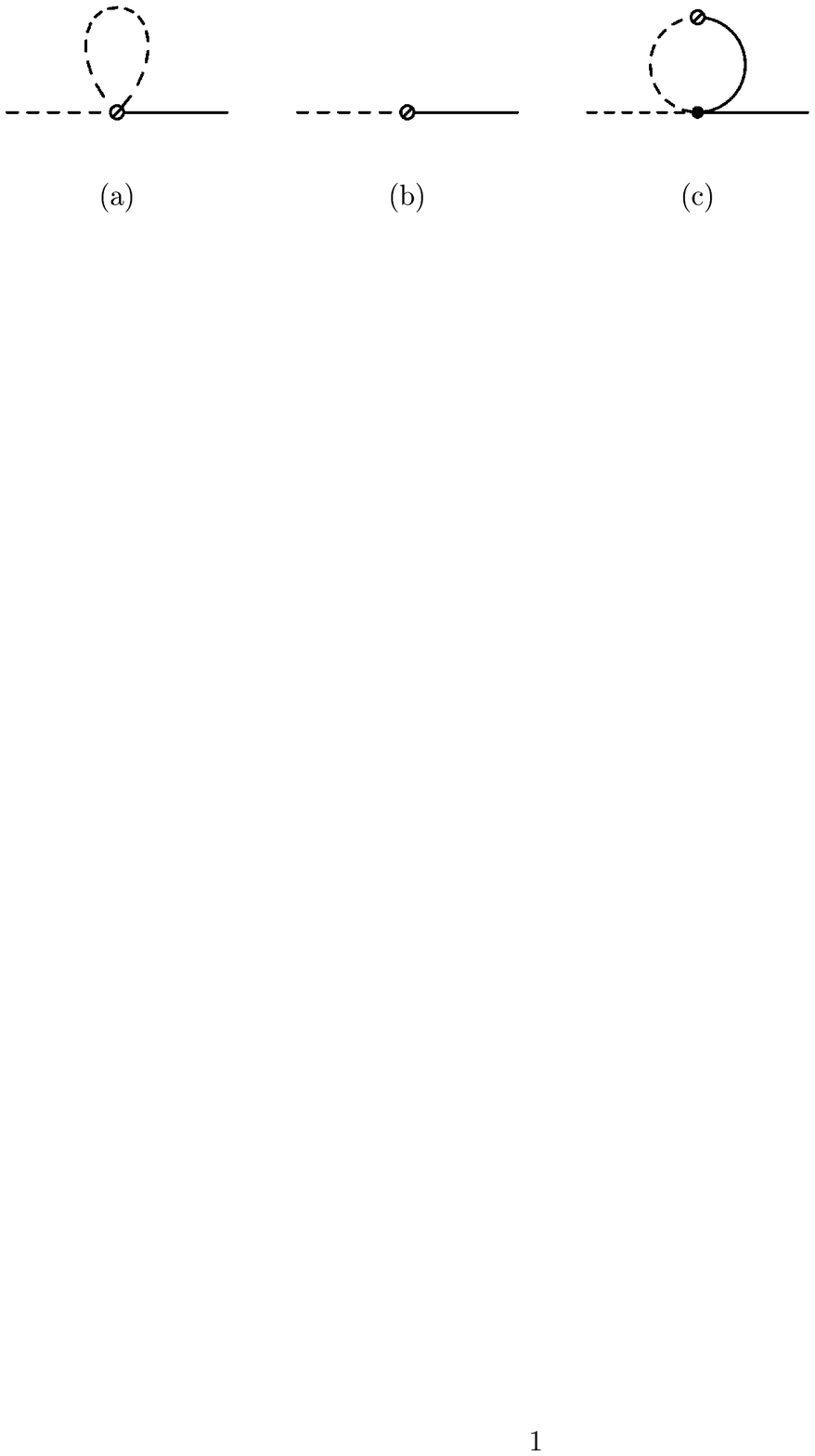}
\caption{Diagrams associated with the cancellation of the $\ord{g}$
violation of the continuum susy Ward identity.  The 2-point and 4-point
shaded vertices that violate
fermion number are those arising from $Q_1 S_c$,
where $S_c$ is the corrected action indicated by \myref{jiil}.
The sum of these diagrams vanishes in the $a \to 0$ limit,
provided the external momentum satisfies $|p_{ext}| \ll a^{-1}$.
\label{fd2}}
}

\subsection{Power counting for the susy lattice theory}
\label{pcsl}
In the transfer matrix analysis of \cite{Giedt:2004qs}, it was
already proven for the susy lattice action \myref{syac}
that the correct continuum limit is
obtained without the need for counterterms.  However, it
is interesting to see the detailed cancellations that bring this
about in perturbation theory.  

In the special case \myref{siex},
the interaction part of the action is given by:
\beq
a^{-1} S_{\stxt{int}}^{\stxt{susy}} 
= \sum_i \( mgx_i^4 + \half g^2 x_i^6 + g x_i^3 \Dels x_i
- \frac{rag}{2} x_i^3 \Delta^2 x_i + 3g x_i^2 \psib_i \psi_i \)~.
\label{slbin}
\eeq
Two new terms appear in \myref{slbin}, relative
to the naive lattice theory \myref{sintn}.  
Both of these vertices scale like $a^{-1}$ and
thus contribute to $D$.  We include these into the 
definition of the 4-point boson vertex, associated with $V_4$
in our power counting analysis.  We then obtain a
slightly modified version of \myref{pcnt}:
\beq
&& D=L+V_4-2I_B-I_F~, \qquad
L=1+I_B+I_F-V_4-\tilde V_4-V_6~, \nnn
&& E_B+2I_B=4V_4+6V_6+2\tilde V_4~, \qquad
E_F+2I_F=2\tilde V_4~.
\eeq
It is straightforward to show that the only solutions
with $D\geq 0$, $L>0$ are just \myref{d0vnt} and
\beq
D=E_F=I_F=\tilde V_4=V_6=0~, \quad L=I_B=V_4=1~, \quad E_B=2~,
\eeq
corresponding to the scalar loop Fig.~\ref{fd1}(b).
Thus in the susy theory we have two $D=0$ 1-loop diagrams
to sum in the proper vertex associated with the boson
2-point function.  It suffices to check at vanishing
external momentum $p_{ext}$, as higher orders in the expansion
about $p_{ext}=0$ will have $D<0$.  One obtains from the boson
loop diagram
\beq
- \frac{6g}{Na} \sum_{k=0}^{N-1} \frac{2m + 2r a^{-1} \sin^2(\pi k/N)}
{a^{-2} \sin^2(2 \pi k/N) + (m + 2r a^{-1} \sin^2(\pi k/N))^2} ~.
\eeq
Added to the fermion loop diagram \myref{latfl}, the
$D=0$ contributions cancel.  This is a consequence of the
1 exact susy of the lattice action.  What is left over is
a $D= -1$ result.  By the theorem of Reisz \cite{Reisz:1987da}, 
as $a \to 0$ this is guaranteed to give the
continuum result---a finite mass correction $-3g$.
(In the present case, it is trivial to check this
explicitly.)

\section{Simulation results}
\label{swna}
All of our simulation results correspond to the
special case of superpotential \myref{siex} and $r=1$.
We will take $m=10$, for the following reason.  
The effects of interactions always cause the mass
gap $m_1$ (energy of the first excitation) to satisfy $m_1 > m$.
Thus the Compton wavelength of this mode is always
shorter than 1/10.  For this reason we can safely
set the system size $\beta=Na$ to $\beta=1$:  the finite
size effects should be negligible since several Compton
wavelengths fit within the total volume, giving
a good approximation to the $\beta \to \infty$ limit.

All of our simulations are performed using hybrid Monte Carlo
techniques.  See \cite{Catterall:2000rv} for details
relating to the implementation for lattice SQM systems.
Unlike \cite{Catterall:2000rv}, we do not use Fourier acceleration; rather,
we have measured autocorrelation times and simply
spaced our samples appropriately.  Due to the simplicity 
of the fermion matrix, we were able to
invert $M^T M s = \phi$ analytically,
where $\phi$ is a pseudofermion configuration, using
two steps of Gaussian elimination.  Alternatively,
since $M^T M$ is a cyclic tridiagonal matrix,
we were able to apply the Sherman-Morrison algorithms with
partial pivoting.  In either case, for the double precision
that we used, we were able to keep the relative residual
error $|(M^T M) \cdot s_{est} - \phi|/|\phi|$ of the inversion to 
less than $1 \times 10^{-14}$ throughout the simulations.
This is far more accurate than what was obtained
using conjugate gradient.  Finally, fit errors due to
excited states were avoided by only fitting times $t$ of
Green functions $G(t)$ such that these effects could be
shown to be negligible.

We have extracted excitation energies, or, effective
masses, from connected Green functions $G^I(t)$
where $t=a, 2a, \ldots, Na$ is the imaginary-time of
points on the lattice, and
$I \in \{ 1B, 1F, 2B, 2F \}$ labels which excitation dominates
the large-time behavior of the Green function.  In particular,
we choose
\beq
&& G^{1B}(t) = \vev{x_1 x_{1+t/a}}~, \qquad
G^{1F}(t) = \vev{\psi_1 \psib_{1+t/a}}~, \nnn
&& G^{2B}(t) = \vev{x_1^2 x_{1+t/a}^2}_{conn.}~, \qquad
G^{2F}(t) = \vev{x_1 \psi_1 x_{1+t/a} \psib_{1+t/a}} ~.
\label{dvevs}
\eeq
Due to the symmetry $x \to -x$ of the action,
as well as fermion number, the states that contribute
to each of these Green functions come from different
sectors of the state space.  For $t \ll Na$ and $N \gg 1$, we have
for example
\beq
G^{1B}(t) = c_{1B} e^{- m_{1B} t} + c_{3B} e^{- m_{3B} t} + \ldots ~,\quad
G^{2B}(t) = c_{2B} e^{- m_{2B} t} + c_{4B} e^{- m_{4B} t} + \ldots ~,
\label{gfex}
\eeq
and similar equations for the fermions.
Here $m_{1B} < m_{2B} < m_{3B} < m_{4B}$.

We now discuss how the contamination of higher
excitations can be suppressed in \myref{gfex}, for
the purpose of fitting the leading mode.
In the limit of $g=0$, we just
have the simple harmonic oscillator with unit mass,
frequency $\omega = m_{1B}$, as well as
fermionic excitations that are degenerate with the bosons.
For $g \not=0$, we expect a spectrum similar to the
anharmonic oscillator, so that $m_{3B}-m_{1B} > 2 m_{1B}$
and similarly for the fermions.  Thus the contribution
of excited states to the Green function $G^{1B}(t)$
can be neglected if $t \gappeq 2/m_{1B}$, since the
relative exponential suppression in \myref{gfex}
will be something smaller than $e^{-4}$.
To avoid finite size effects, which
are of order $\exp[-(Na-t)m_{1B}]$,
we choose $t \lappeq Na/3$.  (Recall that $\beta=Na \equiv 1$
and $m_{1B} > 10$ in our simulations.)
Taking these constraints on $t$ into account, we can safely linearize the fit to
obtain the leading excitation for each Green function (i.e.,
fit $\ln G(t)$ to a linear function in $t$).

Three different versions of the naive action have
been simulated:  $S$, eq.~\myref{naiS}, which has no corrections;
$S_c$, eq.~\myref{jiil}, which has the the $\ord{1}$ counterterm;
and $S_{ca}$, eq.~\myref{impS}, which includes the $\ord{a}$ improvement.

As a first case, we take $g=10$, where the perturbation
expansion parameter $g/m^2=1/10$ is small.  
The results for $m_{1B}$ are shown in Fig.~\ref{fbe}, and for
$m_{1F}$ in Fig.~\ref{ffe}.  We compare these
results to those of the continuum, obtained by
a numerical solution of the Schr\"odinger equation
corresponding to \myref{hjre}.
It can be seen that the uncorrected naive action fails 
to approach the continuum limit for the boson.  The discrepancy
is of order 10 percent, consistent with an $\ord{g/m^2}$ effect.
The $\ord{1}$-subtracted and $\ord{a}$-improved results
are in reasonable agreement with the continuum, though
the latter is clearly better.
The fermion mass does not appear to suffer from the
same error, which is consistent with the fact that
a counterterm is not required for it.  However, it can
be seen that the $\ord{a}$-improved action $S_{ca}$ more rapidly
approaches the continuum result.  (Recall that the correction
to the fermion action occurs only at $\ord{a}$.)
In fact, at the largest
$N$ that we were able to achieve, $S_{ca}$ obviously
gives a superior approximation to the continuum results.
We have also verified that the second set of excitations
$m_{2B}$ and $m_{2F}$ approach the continuum results for
the $\ord{a}$-improved action $S_{ca}$.  For brevity we
do not present these results in detail.

\FIGURE{
\includegraphics[height=6.0in,width=4.0in,angle=90]{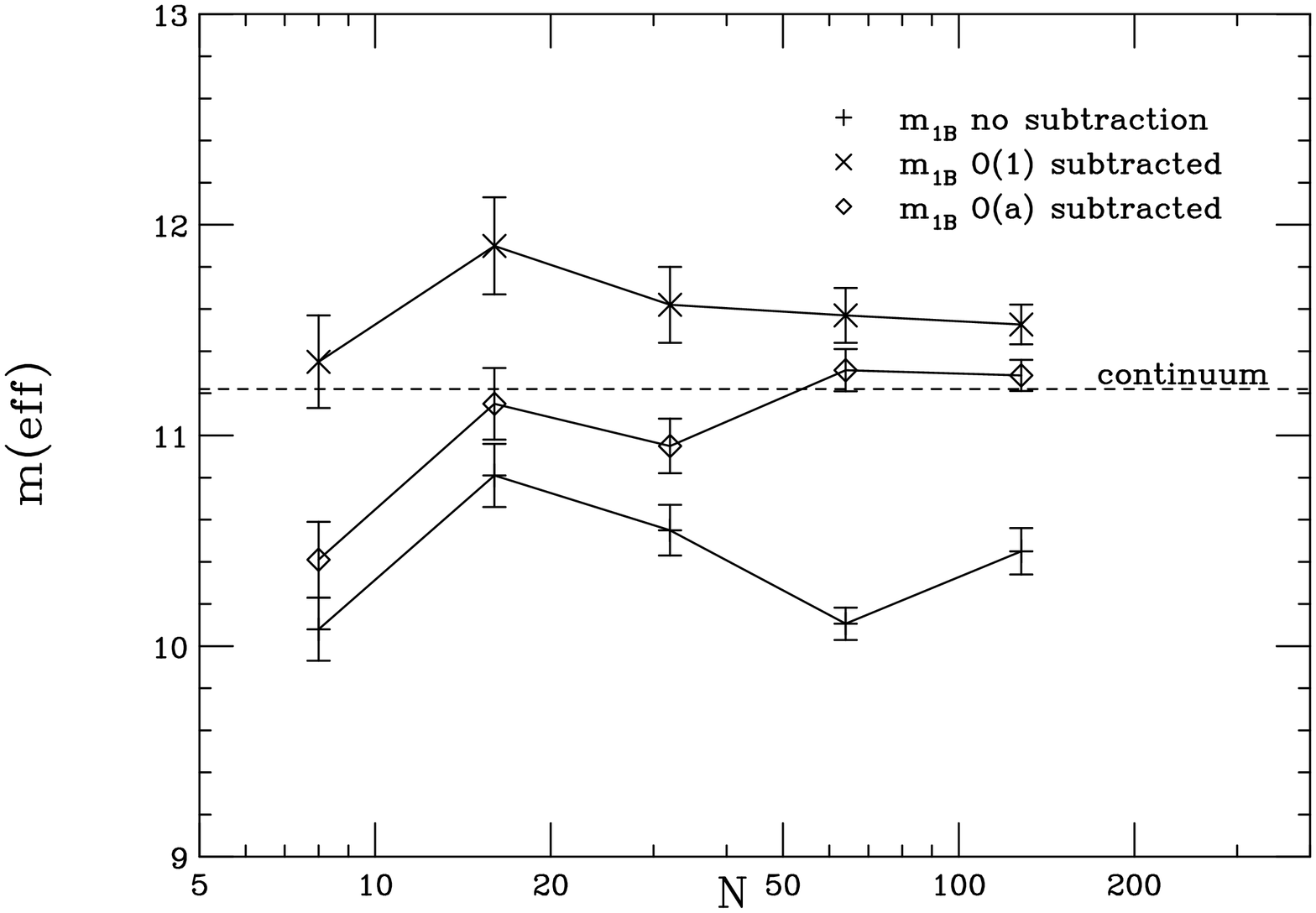}
\caption{Leading boson mass for various forms of the naive action,
with bare parameters $m=g=10$. 
Large $N$ corresponds to the continuum limit with $\beta=Na$
held fixed at $\beta=1$.  Lines are drawn to guide the eye.
\label{fbe}}
}

\FIGURE{
\includegraphics[height=6.0in,width=4.0in,angle=90]{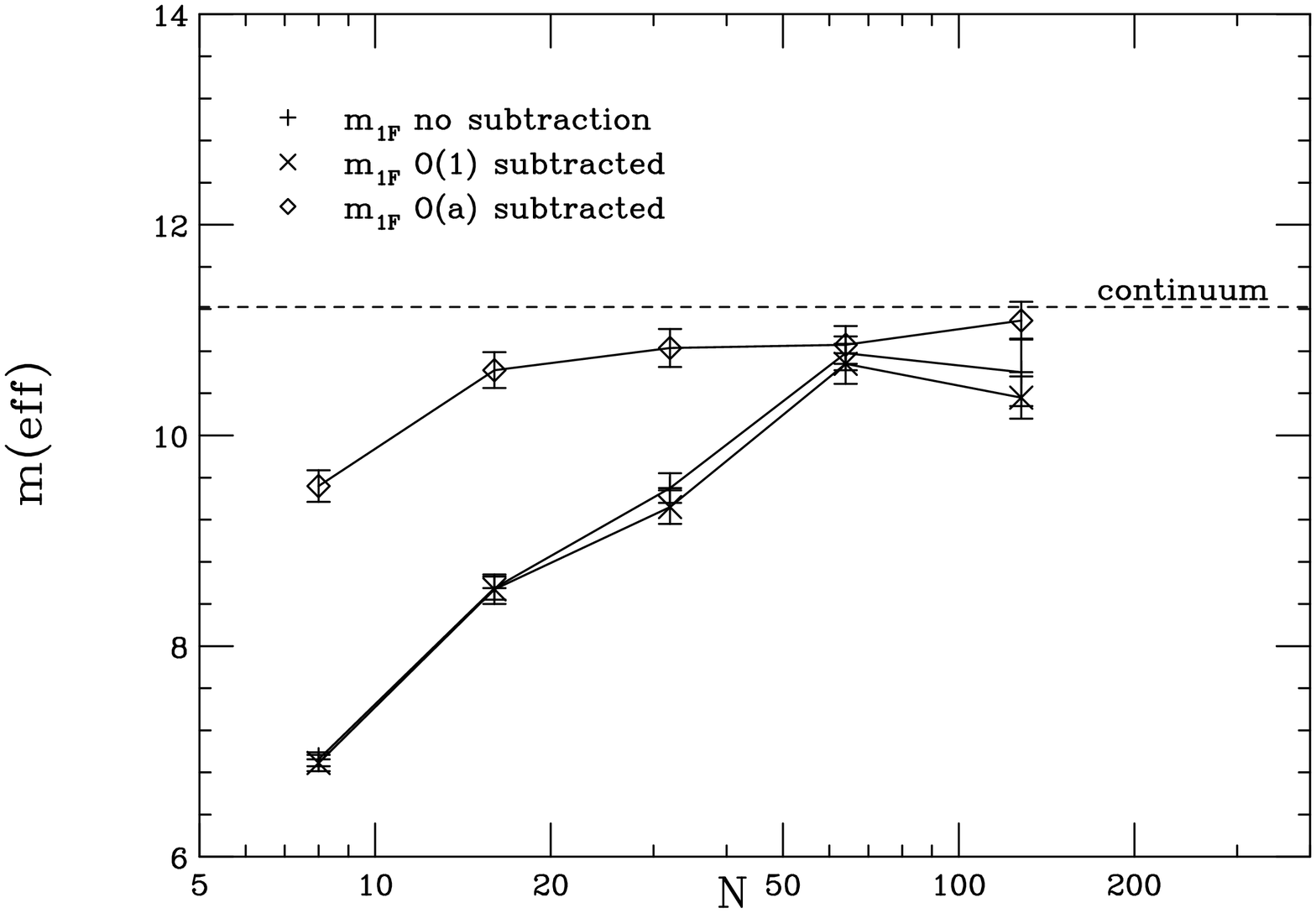}
\caption{Similar to Fig.~\ref{fbe},
except that the fermion mass is displayed. \label{ffe}}
}

Next, we have simulated the $\ord{a}$-improved action
at strong coupling, $g=100$.  The expansion parameter
of perturbation theory would in this case be $g/m^2 = 1$,
clearly beyond the range of validity of a perturbative approach.
In Fig.~\ref{fsc} we give our results for the first
boson and fermion excitations.  It can be seen that,
as indicated by the nonperturbative transfer matrix
analysis of Section \ref{trma}, the desired continuum
limit is obtained.
For comparison, we show the results of CG \cite{Catterall:2000rv},
obtained using the susy lattice action \myref{syac}.
Although the $\ord{a}$-improvement of the naive lattice action gives
a superior approximation to the continuum,
an $\ord{a}$-improvement of the
susy lattice action is straightforward to perform
using the transfer matrix techniques of Section~\ref{trma};
this would also yield better agreement with the continuum.

\FIGURE{
\includegraphics[height=6.0in,width=4.0in,angle=90]{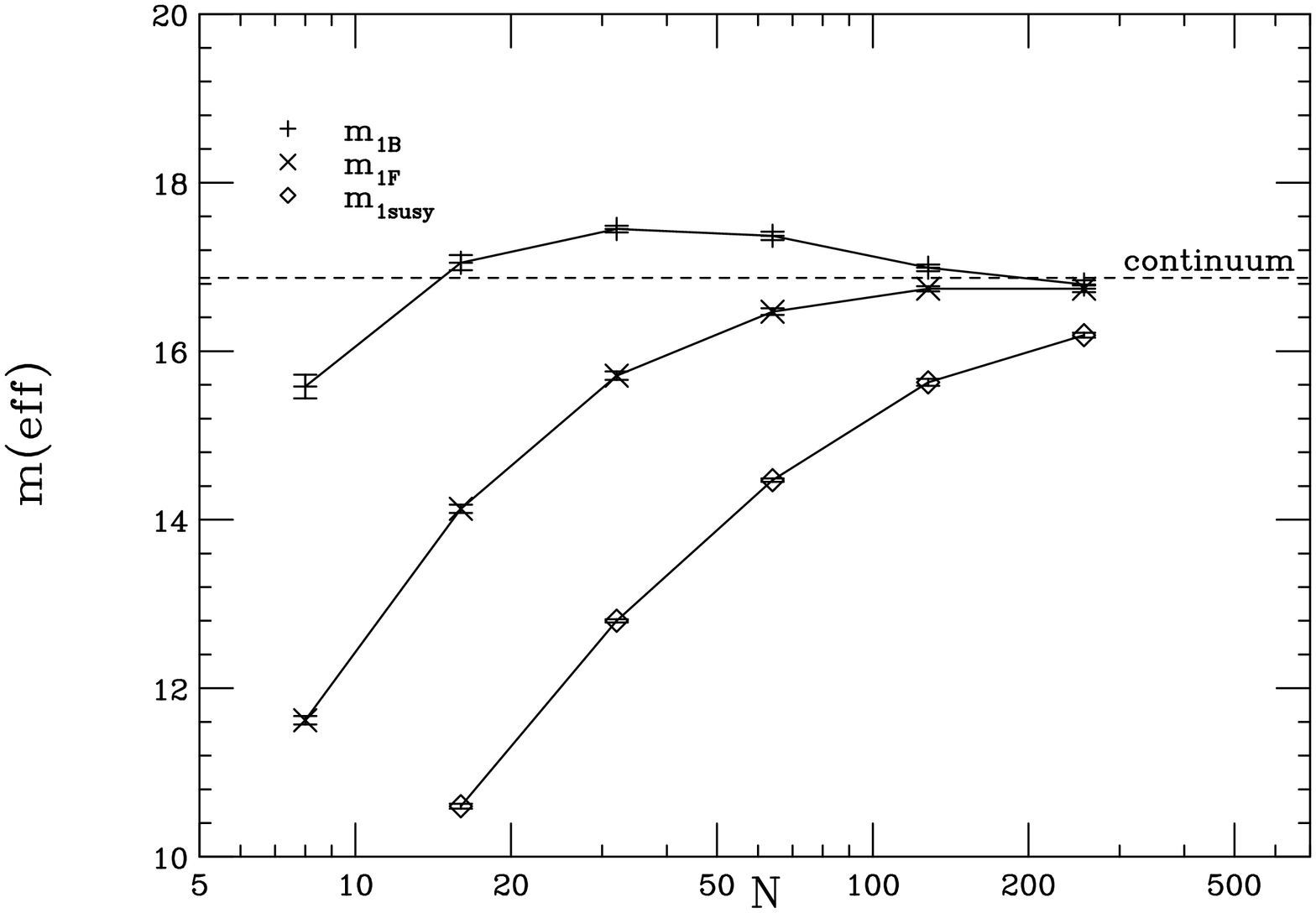}
\caption{Leading boson and fermion mass for strong coupling,
$g=100, m=10$, in the case of the $\ord{a}$-improved naive
action. It can be seen that a good approximation to
the continuum results is obtained for reasonable $N$.
For comparison, we also plot the susy lattice results
reported in \cite{Catterall:2000rv} (the boson and fermion masses
are degenerate).  
\label{fsc}}
}

\section{Conclusions}
\label{concl}
In this article we have shown that a careful study
of lattice power counting, or the transfer matrix,
is necessary in order to
understand the quantum continuum limit, even in the simple
case of 1d SQM.  The lower UV degrees
that occur in lattice perturbation theory with some amount of
exact susy provide a compelling indication of the following:  one should
make exact lattice susy a requirement of the construction
to the extent that
it is possible.  As has been seen, in the case of the
susy lattice action of SQM, the continuum
limit is obtained without the need for counterterms,
due to cancellations between diagrams with $D=0$.
The situation is similar in the field theory context 
(for example, compare to Appendix D of \cite{Giedt:2004qs}); 
although, in field theory we expect
that some amount of fine-tuning of lattice counterterms will generically
be required, particularly if the continuum field theory is not
finite.  This is because modes near the UV
cutoff will not decouple in the $D \geq 0$ diagrams,
yielding contributions that must be subtracted in order
to match the continuum theory.  Nevertheless, we expect
the UV behavior of the lattice theory will be better
if some exact susy is preserved.  Practically speaking,
this amounts to less counterterms that need to be
calculated and adjusted.  This is particularly important
in a strong coupling regime, where the needed fine-tuning of lattice
counterterms must be determined from a nonperturbative
analysis such as Monte Carlo simulation.

It is also of interest to study these issues in models
where symmetries other than susy 
are partly responsible for good UV behavior in
the continuum theory.  For example,
certain classes of nonlinear $\s$-models are known
to be renormalizable in the continuum because the form
of counterterms is greatly restricted by the
nonlinearly realized symmetry.  It has been found that
it is not possible to simultaneously preserve an exact
lattice supersymmetry and the full nonlinear 
symmetry \cite{Giedt:2004qs,Catterall:2003uf}.
Thus we expect the UV behavior of these models to
be worse than in the continuum, and we are presently
studying whether or not a finite set of counterterms
suffices to guarantee the continuum limit.

\vspace{20pt}

\noindent
{\bf \large Acknowledgements}

\vspace{5pt}
We would like to thank Simon Catterall for
useful discussions. This work was supported 
by the National Science and Engineering 
Research Council of Canada and the Ontario 
Premier's Research Excellence Award.

\appendix

\vspace{15pt}
\noindent
{\bf \large Appendix}

\section{The transfer matrix of the naive lattice partition function}
	
The purpose of this appendix is to show that  eqn.~(\ref{Tmatrix})  is the transfer matrix of the 
partition function of the naive lattice action (\ref{naiveaction}), with Wilson parameter $r = 1$  (see (\ref{wofr})), supplemented for further use below by the counterterms (\ref{cterms1}):
\beq
S[k, \ell]  &=&   \sum\limits_{i=1}^N \[  {1 \over 2a} (x_i - x_{i-1})^2 
+ {1 \over 2a}( h'_i h'_i +   k_i )+ \psib_i (\psi_{i} - \psi_{i-1} )
+  a  (h''_i + \ell_i) \psib_i \psi_i \] ~, 
\label{Skel}
\eeq
where, as usual $k_i = k(x_i), \ell_i = \ell(x_i)$, 
with $i+N \equiv i$. 
 The lattice partition function is then defined as:
\beq
\label{QMZ}
Z[k, \ell] = c \prod\limits_{i=1}^N \int d\bar\psi^i  d \psi^i \int\limits_{- \infty}^{\infty} d x^i \;  e^{- S}~. 
\eeq 
To continue, it is convenient to change the fermionic variables as follows: 
\beq
\label{change}
\psi_i = \bar\eta_{i+1}, ~ \psib_i = \eta_i,
\eeq
so that the action (\ref{Skel}) and partition function (\ref{QMZ}) become:
\beq
\label{change2}
S[k, \ell]  &=& \sum\limits_{i=1}^N  \[ {1 \over 2a} (x_i - x_{i-1})^2 
+ {1 \over 2a}( h'_i h'_i +   k_i ) + \bar\eta_i \eta_i -  
 \bar\eta_{i+1} \eta_i \left(1+  a  (h''_i + \ell_i) \right) \] , \nonumber \\
Z[k, \ell] &=& 
\tilde{c}  \prod\limits_{i=1}^N \int d\bar\eta^i  d \eta^i \int\limits_{- \infty}^{\infty} d x^i \;  e^{- S}~,
\eeq 
where we absorbed a minus sign in the normalization constant $\tilde{c}$. 

To construct the transfer matrix and hamiltonian (see, for example, \cite{Montvay}), we first introduce, at each time slice, a Hilbert space which is  a tensor product of a bosonic and fermionic space. The bosonic Hilbert space is that of square integrable functions on the line. We use the  basis of position eigenstates, $\{\vert x \rangle, \; \langle x^\prime | x \rangle = \delta(x^\prime - x) \}$,   where the momentum and position operators, $\[ \hat{p}, \hat{q} \] = - i$, act as $\hat{q} \vert x \rangle = \vert x \rangle x$ and $e^{i \hat{p} \Delta } \vert x \rangle= \vert x + \Delta \rangle$ (note that we continue using the dimensions of the previous section: $x$ has mass dimension $-1/2$, $a$ has dimension of length, while the superpotential $h(x)$ is dimensionless). The fermionic Hilbert space is two dimensional and is  spanned by the vectors $\vert 0 \rangle$ and $\vert 1 \rangle$. The  fermionic creation and annihilation operators obey $\{\hat{b}^\dagger, \hat{b} \}= 1$, such that $\hat{b} \vert 0 \rangle = 0$, $\vert 1 \rangle = \hat{b}^\dagger \vert 0 \rangle$. 
The fermionic coherent states are defined as $\vert \eta \rangle\equiv \vert 0 \rangle + \vert 1 \rangle  \eta$, $\langle \eta \vert = \langle 0 \vert + \bar\eta \langle 1 \vert$, where $\eta$ and $\bar\eta$ are Grassmann variables. We then recall the usual relations for the decomposition of unity, $\langle \eta^\prime \vert \eta \rangle = e^{\bar\eta^\prime \eta}$; ${\hat{1}} = \int  d\bar\eta d \eta e^{- \bar\eta\eta} \vert \eta \rangle \langle \eta \vert$, and  for traces of operators ${\cal{O}}$ on the fermionic Hilbert space: 
Tr$\; {\cal{O}} = \int  d\bar\eta d \eta e^{- \bar\eta\eta} \langle \eta\vert {\cal{O}} \vert - \eta \rangle$;   Tr$(-1)^F {\cal{O}} = \int  d\bar\eta d \eta e^{- \bar\eta\eta} \langle \eta\vert {\cal{O}} \vert  \eta \rangle$, with  $(-1)^F \vert 0 \rangle = \vert 0 \rangle$.
We then define the transfer matrix  by the equality:
\beq
\label{aTmatrix}
Z[k,\ell] &=& \tilde{c} \prod\limits_{i=1}^N \int d\bar\eta^i  d \eta^i \int\limits_{- \infty}^{\infty} d x^i \;  e^{- S} \equiv {\rm Tr} \; (-1)^F \; \hat{T}^N   = \\
&=&  \prod\limits_{i=1}^N \int d\bar\eta^i  d \eta^i e^{- \bar\eta^i \eta^i} \int\limits_{- \infty}^{\infty} d x^i \times \nonumber \\
&\times& \langle \eta^N, x^N \vert \hat{T} \vert \eta^{N-1}, x^{N-1} \rangle \langle \eta^{N-1}, x^{N-1} \vert \hat{T} \vert \eta^{N-2}, 
x^{N-2} \rangle \times \ldots \nnn
&& \times \langle \eta^2, x^2 \vert \hat{T} \vert \eta^{1}, 
x^{1} \rangle \langle \eta^1, x^1 \vert \hat{T} \vert \eta^{N}, x^{N} \rangle \nonumber~,
\eeq
or, equivalently, through its matrix elements,  $\langle \eta^{i+1}, x^{i+1} \vert \hat{T} \vert \eta^{i}, x^{i} \rangle =$ 
\beq
\label{Tmatrix2}
\tilde{c}^{1/N} \;  \exp \left[-  {a\over 2}  \left(  x^{i+1} -  x^{i} \over a \right)^2  -  {a\over 2}( (h^\prime (x^{i+1})^2 + k(x^i)) +  \bar\eta^{i+1}  \eta^i 
 \left( 1+ a (h^{\prime \prime} (x^{i}) + \ell(x^i))   \right)  \right].
\eeq
(Substituting (\ref{Tmatrix2}) into (\ref{aTmatrix}) one can immediately obtain (\ref{change2})). 

Now, we can use the fermion coherent state identity (following from the definitions given above)  $\langle \eta^\prime | 1 - X \hat{b}^\dagger \hat{b} |\eta    \rangle = e^{(1-X) \bar\eta^\prime \eta}$, as well as the action of $\hat{p}, \hat{q}$ on the states $|x^i\rangle$ , to verify that the transfer matrix operator
 $\hat{T}$, with matrix elements   given by eqn.~(\ref{Tmatrix2}), is: 
\beq
\label{Tmatrix3}
\hat{T} \; = \; \tilde{c}^{1/N} \int\limits_{-\infty}^\infty d z\; \exp\left( - {z^2 \over 2 a}  - {a \over 2} (h^\prime (\hat{q})^2 + k(\hat{q}))  \right)  \; \exp \left(i z \hat{p} \right)\;  \left(1 + a (h^{\prime\prime}(\hat{q}) + \ell(\hat{q}))\; \hat{b}^\dagger \hat{b} \right)~,
\eeq
as already stated in eqn.~(\ref{Tmatrix}) in the main text (where $\tilde{c}^{1/N}$ is called ${\cal{N}}(a)$). 

We note that the properties of the lattice partition function and transfer matrix of the supersymmetric lattice action are studied in \cite{Giedt:2004qs}.

\end{document}